\documentclass[a4paper,11pt]{article}
\usepackage{pos}
\usepackage{wrapfig}
\usepackage{setspace}
\usepackage{cleveref}

\title{Advances in lattice hadron physics calculations using the gradient flow}

\author*[a,1]{K.U.~Can}
\author[b]{R.~Horsley}
\author[c]{Y.~Nakamura}
\author[d]{H.~Perlt}
\author[e]{P.E.L.~Rakow}
\author[f]{G.~Schierholz}
\author[g]{H.~St\"{u}ben}
\author[a]{R.D.~Young}
\author*[a,1]{J.M.~Zanotti}

\affiliation[a]{CSSM, Department of Physics, The University of Adelaide, 
Adelaide SA 5005, Australia.}
\affiliation[b]{School of Physics and Astronomy, University of Edinburgh,
Edinburgh EH9 3JZ, UK.}
\affiliation[c]{RIKEN Center for Computational Science,
Kobe, Hyogo 650-0047, Japan.}
\affiliation[d]{Institut f\"{u}r Theoretische Physik, Universit\"{a}t Leipzig,
04103 Leipzig, Germany.}
\affiliation[e]{Theoretical Physics Division, Department of Mathematical Sciences, University of Liverpool,\\ Liverpool L69 3BX, UK.}
\affiliation[f]{Deutsches Elektronen-Synchrotron DESY, 
Notkestr. 85, 22607 Hamburg, Germany.}
\affiliation[f]{Regionales Rechenzentrum, Universit\"{a}t Hamburg,
20146 Hamburg, Germany.}

\emailAdd{kadirutku.can@adelaide.edu.au}
\emailAdd{james.zanotti@adelaide.edu.au}

\note{For the QCDSF-UKQCD-CSSM Collaborations.}

\abstract{Lattice calculations of hadronic observables are aggravated by short-distance fluctuations. The gradient flow, which can be viewed as a particular realisation of the coarse-graining step of momentum space RG transformations, proves a powerful tool for evolving the lattice gauge field to successively longer length scales for any initial coupling. Already at small flow times we find the signal-to-noise ratio of two- and three-point functions significantly enhanced and the projection onto the ground state largely improved, while the effect on the hadronic observables considered here to be negligible. A further benefit is that far fewer conjugate gradient iterations are needed for the Wilson-Dirac inverter to converge. Additionally, we find the renormalisation constants of quark bilinears to be significantly closer to unity.}

\FullConference{%
 The 38th International Symposium on Lattice Field Theory, LATTICE2021
  26th-30th July, 2021
  Zoom/Gather@Massachusetts Institute of Technology
}

\begin{document}
\maketitle

\section{Introduction}
The gradient flow approach has proven itself to be a powerful tool in investigating the long distance behaviour of lattice gauge field theories (see for instance the reviews~\cite{Ramos:2015dla,Shindler:2013bia}). In this approach, the fundamental gauge fields, $A_\mu$, are evolved with respect to a fictitious flow time, $\tau > 0$, according to~\cite{Luscher:2010iy},
\begin{align}
    \partial_\tau B_\mu &= D_\nu G_{\nu\mu}, \; \text{where,} 
    \left. B_\mu \right|_{\tau=0} = A_\mu, \; \text{and} \\
    \; G_{\mu\nu} &= \partial_\mu B_\nu - \partial_\nu B_\mu + [B_\mu,B_\nu], \; D_\mu = \partial_\mu + [B_\mu, \cdot].
\end{align}   
The flowing procedure can be viewed as a particular realisation of the coarse-graining step of momentum space renormalisation group transformations~\cite{Abe:2018zdc,Carosso:2018bmz,Luscher:2013vga,Makino:2018rys,Peterson:2021lvb}, which evolves the lattice gauge fields to successively longer length scales for any initial coupling. Flowed fields are smooth renormalised fields such that the correlators of the flowed fields are automatically renormalised~\cite{Luscher:2011bx}. As relevant to this work, this means the renormalisation of quark bilinear operators is expected to become trivial as one flows to larger times, while the renormalised quantities (e.g. quark and hadron masses) remain constant. In the rest of this contribution, we report on the advances of the QCDSF/UKQCD Collaboration's application of the gradient flow approach to extract some select hadronic observables.     

\section{Simulation details and performance}
We perform exploratory calculations on the $\beta=5.5$, $32^3 \times 64$, $2+1$-flavour gauge configurations generated by the QCDSF/UKQCD collaboration. A non-perturbatively $\mathcal{O}(a)$-improved Wilson (Clover) action is used for simulating the dynamical quarks~\cite{Cundy:2009yy}. The lattice spacing of this ensemble is $a=0.074(2) \, {\rm fm}$~\cite{Bornyakov:2015eaa} and the quark masses are tuned to the $SU(3)$ symmetric point, yielding $a M_\pi \simeq 0.175$ in lattice units and $M_\pi \simeq 470 \, {\rm MeV}$ in physical units. More details on this single set of configurations can be found in~\cite{Bietenholz:2010jr,Bietenholz:2011qq}.

In performing calculations, we first flow the original configurations to fixed flow times, and then compute the valence quark propagators which are then used to calculate standard nucleon two- and three-point correlation functions. Considering that the improvement coefficients are expected to be driven to their tree-level improved values as a result of flowing, we employ a tree-level Clover action (i.e. $c_{sw}=1$) in computing the valence quark propagators. A mandatory step is retuning the $\kappa_{val}(\tau)$ at each flow time (including $\tau=0$ since the action parameters have changed from those used in the ensemble generation) to ensure that we are working at fixed quark mass. On each set of (un)flowed configurations we tune $\kappa_{val}(\tau)$ to keep $a M_\pi \simeq 0.175$ consistent with its unitary value. Note that this is a partially-quenched procedure since we do not modify the sea quarks. A similar approach is taken in Refs.~\cite{Berkowitz:2017opd,Miller:2020evg}. If the physics is unaffected by the flowing procedure, one of the quantities that should remain constant with respect to flow time is the $M_\pi/M_N$ ratio. Once $\kappa_{val}(\tau)$ is tuned, we find $M_\pi/M_N \sim 0.38$ is constant in $\tau$, in very good agreement with its value obtained in the unitary simulation~\cite{Bietenholz:2011qq}.  

In the results presented here, we perform calculations on 100 configurations only. Source and sink nucleon interpolating operators are smeared in a gauge-invariant manner by Jacobi smearing. We report only statistical errors in this work which are determined by a bootstrap analysis.
\begin{figure}[ht]
    \centering
    \includegraphics[width=.55\textwidth]{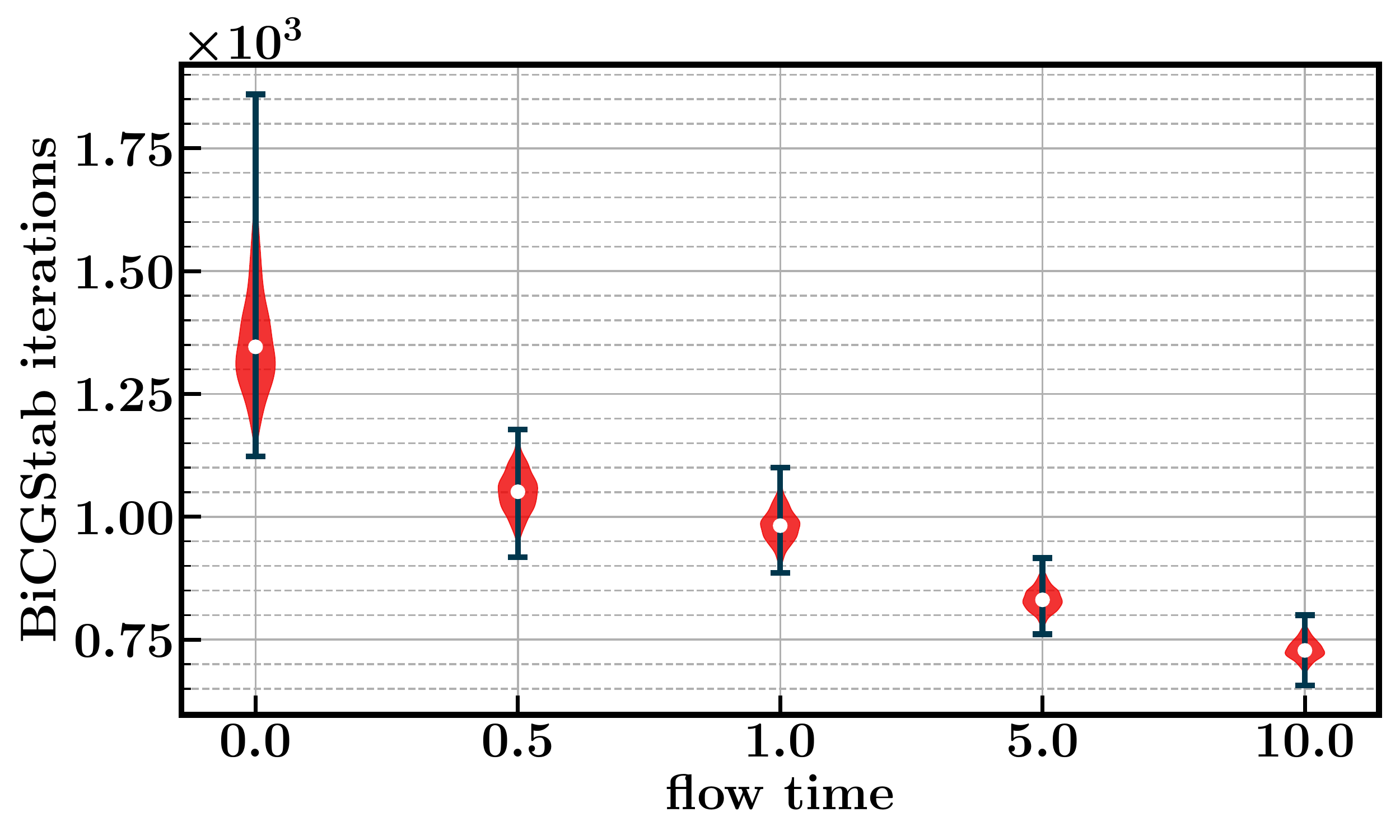}
    \caption{\label{fig:perf}Distribution of the number of iterations (red blobs) it took for the solver to converge at each flow time. The white dots show the means of the distributions. Note that the vertical lines with caps show the full spread of the distributions, not their $1\sigma$ interval.}
\end{figure}

A significant benefit of using flowed configurations is the improvement of the solver performance. We show in \Cref{fig:perf} that the number of iterations required for the Wilson-Dirac solver to converge decrease, along with the number of exceptional configurations, with increasing flow time.

\begin{figure}[ht]
    \centering
    \includegraphics[width=.5\textwidth]{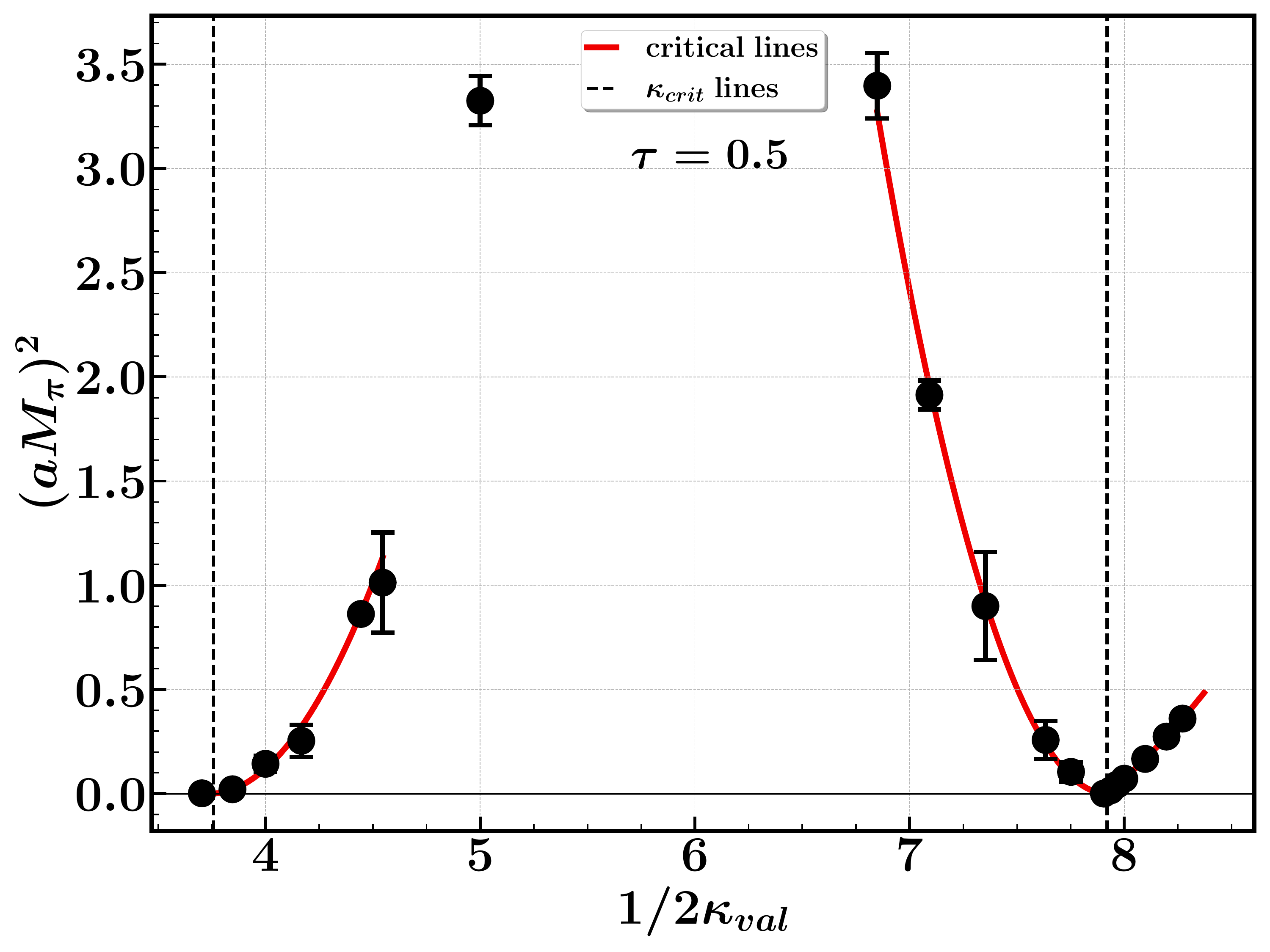}
    \caption{\label{fig:critlines}The structure of the critical lines at $\tau = 0.5$. Black points are $(aM_\pi)^2$ obtained at each $\kappa_{val}$. Red lines are the fits of the form $f((aM_\pi)^2) = b (1/\kappa_{val} - 1/\kappa_{crit}) + c (1/\kappa_{val} - 1/\kappa_{crit})^2$, where $b,c$ are the fit parameters. Vertical dashed lines mark the positions of $\kappa_{crit}$.} 
\end{figure}

A final remark is on the possibility of exploring the Dashen/Aoki phase which has implications for the strong-$CP$ problem and the $U(1)$ anomaly~\cite{PhysRevD.3.1879,Aoki:1983qi,PhysRevLett.57.3136,Sharpe:1998xm,PhysRevD.90.094508}. The improved performance of the solver allows us to compute valence quark propagators for $\kappa_{val}(\tau) \gtrsim \kappa_{crit}(\tau)$ at a reduced computational cost for a range of $\kappa_{val}(\tau)$ values for $\tau > 0$. In our exploratory calculation, we are able to map the phase boundaries (the critical lines where the pion mass vanishes), which separate the normal and $CP$-violating Dashen/Aoki phases for Wilson quarks, as shown in \Cref{fig:critlines}. This structure of the critical lines was confirmed in the quenched approximation before~\cite{Aoki:1996af}. As $\tau \to \infty$, critical points appear at $\kappa_{crit} \equiv 2\kappa_{val} = 1/8, 1/4$. We see that our $\kappa_{crit}$ values appear close to their expected values at finite flow time. Exploration of the Dashen/Aoki phase, however, requires a full-QCD simulation with non degenerate up and down quarks~\cite{PhysRevLett.112.141603}, which we leave to a future study.

\section{Hadronic observables}
\begin{figure}[ht]
    \centering
    \includegraphics[width=.65\textwidth]{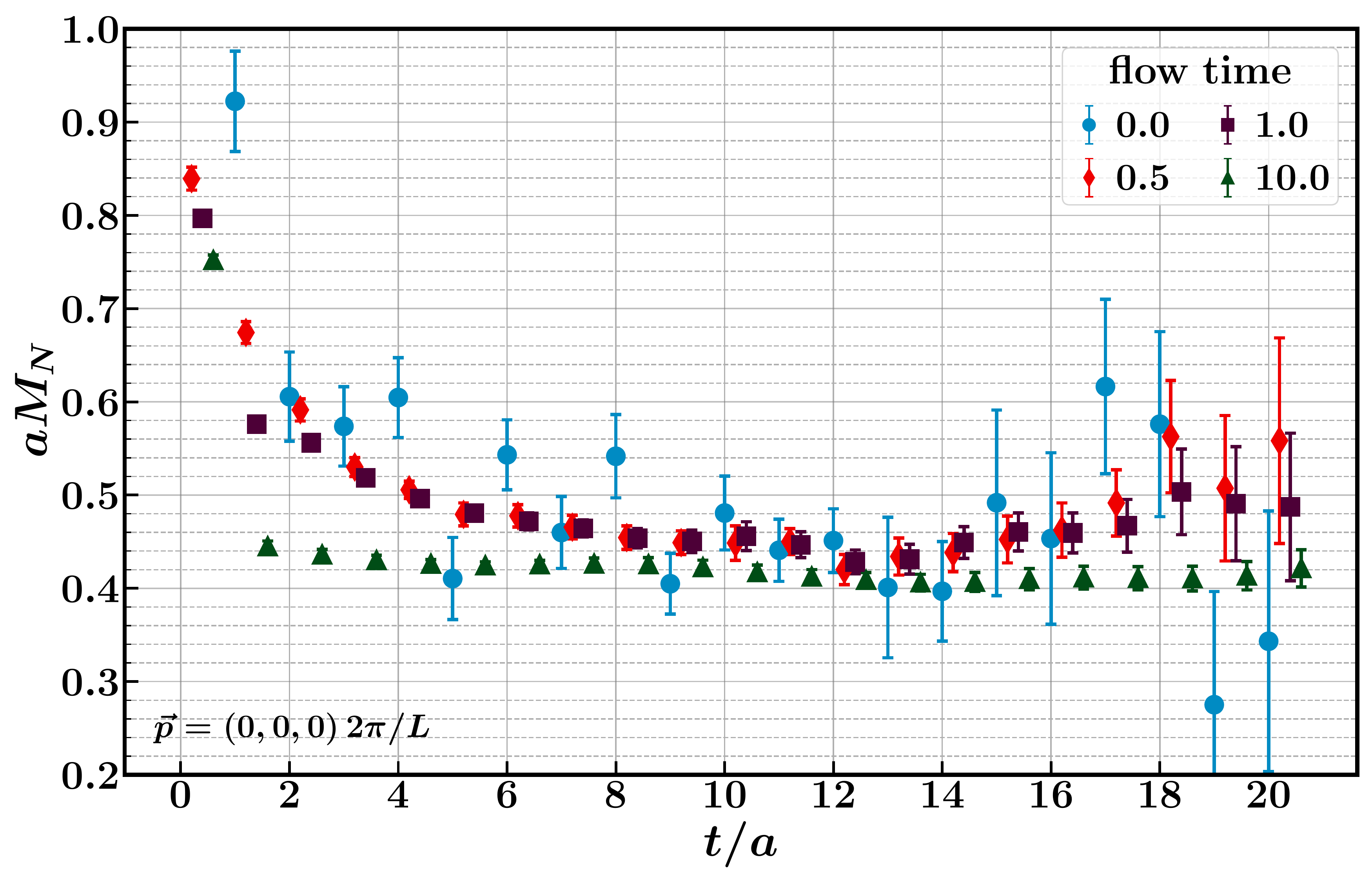} \\
    \includegraphics[width=.65\textwidth]{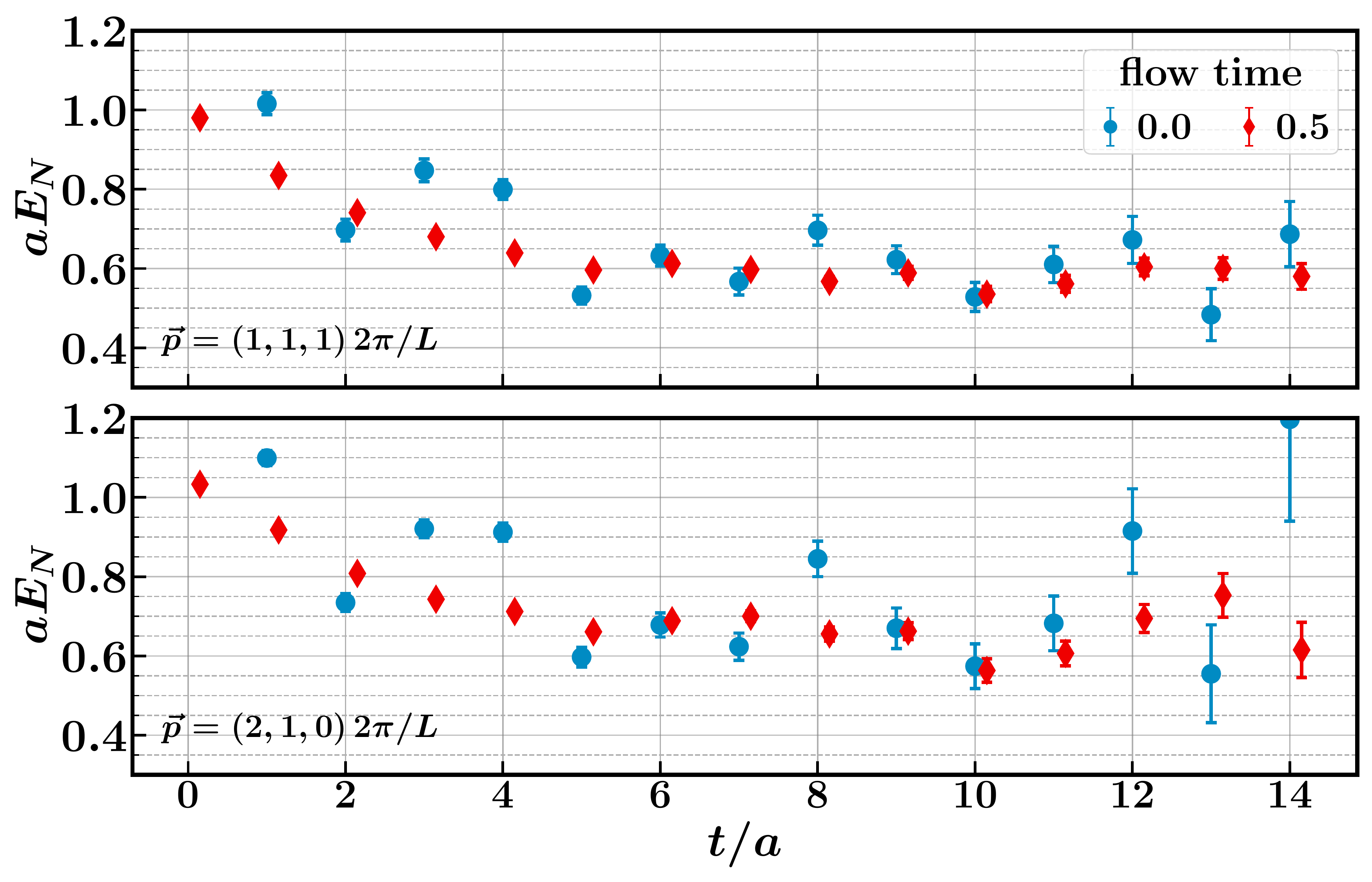}
    \caption{\label{fig:nucl_2pt}Nucleon effective mass plots at varying momenta, comparing the signal for different flow times.}
\end{figure}

In \Cref{fig:nucl_2pt}, we give effective energy plots of a nucleon two-point correlator obtained at different flow times and for a couple of Fourier momenta. There is a clear improvement in the signal quality even with a short amount of flow. Fluctuations in the earlier time slices arising from excited states are tamed, while the signal deterioration from statistical noise is pushed back to later times. We obtain similar nucleon masses via plateau fits at flow times $\tau = 0, 0.5,$ and $1.0$. The mass obtained at $\tau=10.0$, however, is inconsistent with the others, indicating that possibly too much short-distance physics has been removed from the gauge field configurations. Although short flow times seem adequate for improving the quality of the hadronic observables, we leave investigating the effects of longer flows times to a later work. 
\begin{figure}[h]
    \centering
    \includegraphics[width=.69\textwidth]{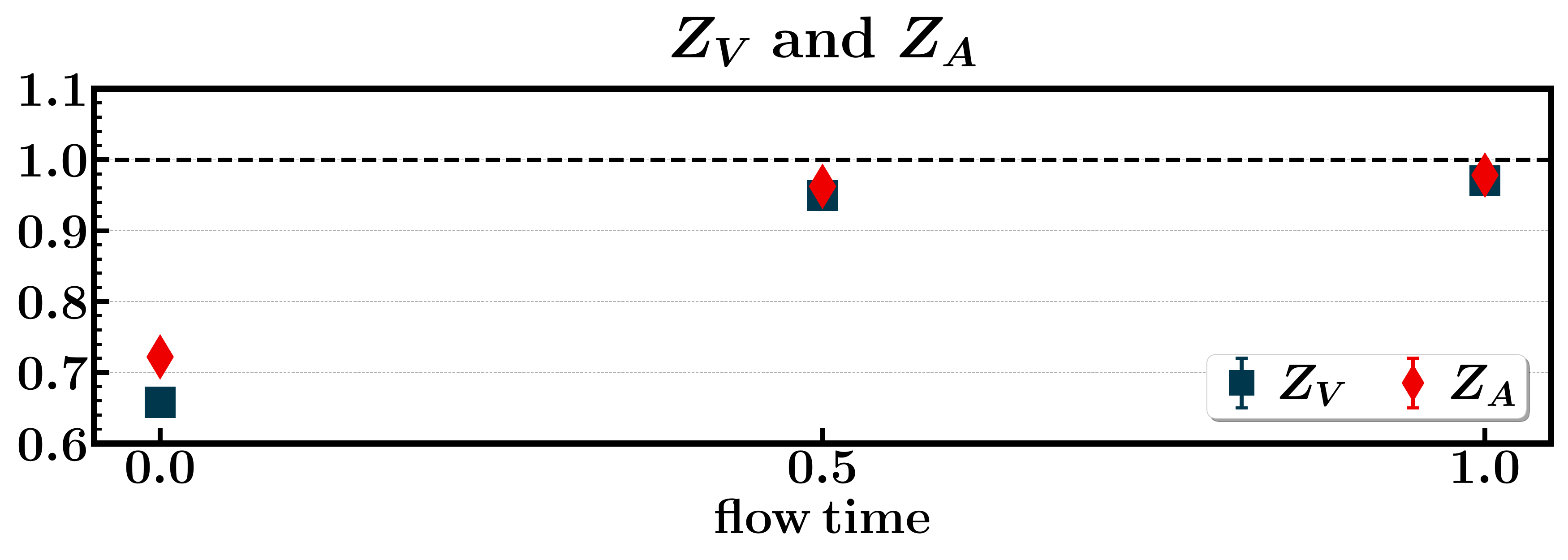}
    \caption{\label{fig:renorm}Vector ($Z_V$) and axial-vector ($Z_A$) renormalisation constants with respect to flow time.}
\end{figure}

As the gauge fields are flowed, one expects the renormalisation of quark bilinears to become trivial, i.e. renormalisation constants of quark bilinears should evolve towards 1 with increasing flow time. We show the vector and axial-vector renormalisation constants calculated at each flow time using the RI${}^\prime$-MOM scheme~\cite{Constantinou:2014fka} in \Cref{fig:renorm} as an example. Their tendency towards unitarity is evident. At the same time, we see that $Z_A/Z_V \to 1$, a feature typically reserved for fermion actions with good chiral symmetry.   

One particular hadronic observable of great interest is the axial charge of the nucleon, whose extraction is prone to systematic effects. One typically needs to control the excited state contamination by maximising the source-sink separation for a reliable, high-precision determination. Controlling these systematics, which have been a highlight of recent calculations~\cite{Djukanovic:2021lat}, demands great computational effort and complicated analysis methods. To address these issues, we investigate the viability of the gradient flow approach as a means of extracting nucleon charges and form factors. 

\begin{figure}[ht]
    \centering
    \includegraphics[width=.7\textwidth]{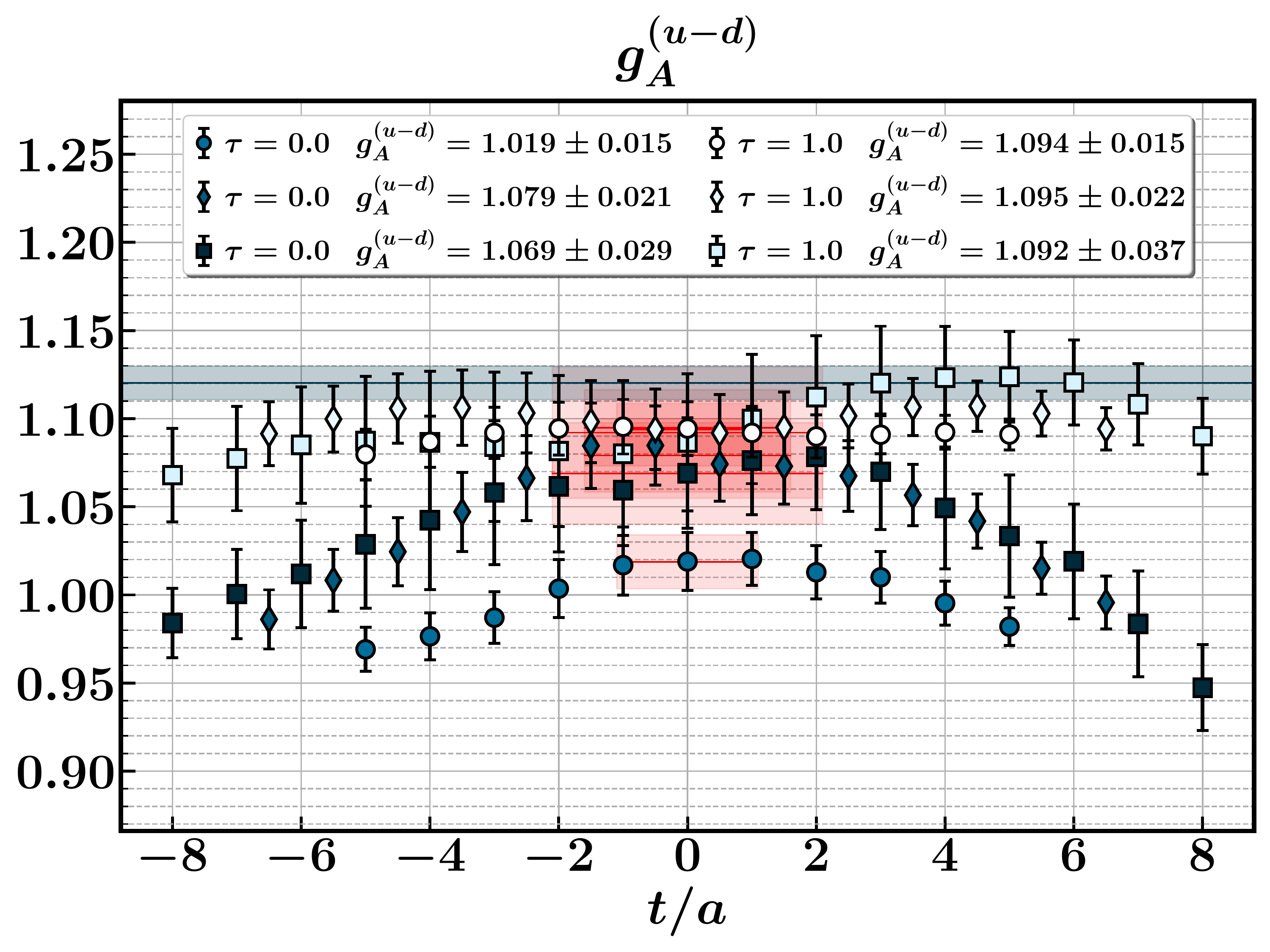} \\
    \caption{\label{fig:ga1}Nucleon isovector axial charge $g_A^{u-d}$ compared to the value extracted (blue bands) from unitary calculation using a variational approach with the full statistics~\cite{Dragos:2016rtx}. Points are centred for a clear comparison of the signals of different source-sink separations. Dark (light) coloured symbols indicate flow time $\tau = 0.0$ $(\tau = 1.0)$ results for $t_{sink}=\{10a \, (\circ), 13a \, (\diamond), 16a \, (\square) \}$.}
\end{figure}

\begin{figure}[ht]
    \centering    
    \includegraphics[width=.7\textwidth]{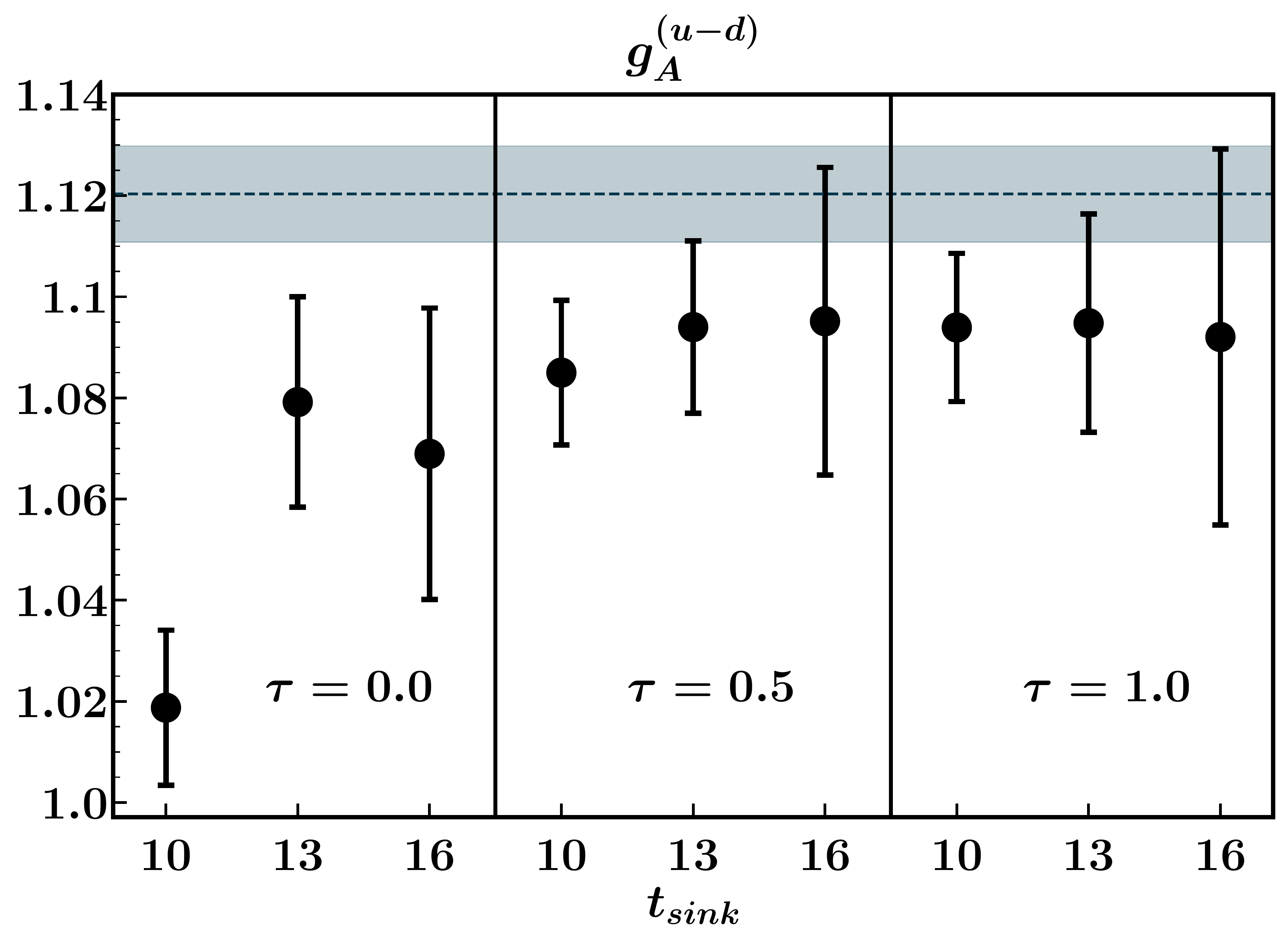}
    \caption{\label{fig:ga2}Summary plot for the nucleon isovector axial charge $g_A^{u-d}$ extracted at each $\tau$ and $t_{sink}$ compared to the value extracted (blue band) from unitary calculation using a variational approach with the full statistics~\cite{Dragos:2016rtx}.}
\end{figure}

We illustrate the computation of $g_A$ in \Cref{fig:ga1,fig:ga2} as an example. 
In \Cref{fig:ga1} we show the signal for the (renormalised) isovector $g_A^{u-d}$ as obtained from a standard three-point function analysis following Ref.~\cite{Dragos:2016rtx}. We calculate the three-point function with a single nucleon interpolating operator for three source-sink separations, $t_{sink}=\{10a, 13a, 16a\}$, at flow times $\tau = 0$ and $1.0$. The fit regions are indicated by red bands and the extracted values are given in the figure legend. There is a striking contrast between the unflowed and flowed results where there is the curvature due to excited states and the familiar source-sink separation dependence in the unflowed case, while the points obtained on the flowed ensemble are almost on top of each other and the curvature is flattened. 

We provide a summary plot of the extracted $g_A^{u-d}$ values in \Cref{fig:ga2}, which includes additional results obtained at $\tau=0.5$ following the same procedure. Finally, both the signal and the extracted values are compared to a previous determination of $g_A$ (blue bands on both figures) on the full original ensemble from a variational analysis~\cite{Dragos:2016rtx}. Note that our results are from 100 measurements only while~\cite{Dragos:2016rtx} have $\mathcal{O}(10^3)$ measurements. Given the difference in statistics, our results are in excellent agreement. The stability of the central value with respect to source-sink separation at $\tau=1.0$ is encouraging for future precision calculations where one can potentially use the shorter separation which has better statistical accuracy. We are currently in the process of repeating this analysis with increased statistics where the results will be reported in a future paper along with other hadronic observables.    

\section{Conclusions}
We have reported on QCDSF/UKQCD Collaboration's advances in extracting several hadronic observables by an application of the gradient flow method. We have shown that working on flowed configurations improves the solver performance and reduces the computation time. Based on the presented results, we have argued that keeping the $a M_\pi$ constant with respect to flow time by retuning the bare quark masses on flowed configurations is enough to keep the physics constant (e.g. $M_\pi / M_N$ remains the same), however a rigorous check is desirable for confirmation. The advantage of the gradient flow is most evident for the hadronic observables where the excited states contamination is tamed and a better signal is obtained for both two- and three-point correlation functions and related quantities. These preliminary results were extracted from a low statistics run. We are working towards a high-statistics calculation.

\acknowledgments
The numerical configuration generation (using the BQCD lattice QCD program~\cite{Haar:2017ubh})) and data analysis (using the Chroma software library~\cite{Edwards:2004sx}) was carried out on the DiRAC Blue Gene Q and Extreme Scaling (EPCC, Edinburgh, UK) and Data Intensive (Cambridge, UK) services, the GCS supercomputers JUQUEEN and JUWELS (NIC, Jülich, Germany) and resources provided by HLRN (The North-German Supercomputer Alliance), the NCI National Facility in Canberra, Australia (supported by the Australian Commonwealth Government) and the Phoenix HPC service (University of Adelaide). RH is supported by STFC through grant ST/P000630/1. HP is supported by DFG Grant No. PE 2792/2-1. PELR is supported in part by the STFC under contract ST/G00062X/1. GS is supported by DFG Grant No. SCHI 179/8-1. KUC, RDY and JMZ are supported by the Australian Research Council grant DP190100297.

\providecommand{\href}[2]{#2}\begingroup
\renewcommand{\baselinestretch}{1}
\setlength{\bibsep}{.25pt}
\setstretch{1}
\raggedright
\endgroup

\end{document}